\documentstyle[preprint,aps]{revtex}

\begin{document}

\title{The System of Multi Color-flux-tubes 
\\ in the Dual Ginzburg-Landau Theory} 
\vspace{3.5cm}
\author{ 
   H. Ichie,  H. Suganuma and H. Toki
}
\address{
           Research Center for Nuclear Physics (RCNP), 
           Osaka University, \\  
       Ibaraki, Osaka 567, Japan }
\maketitle
\vspace{3cm}

\begin{abstract}
We study the system of multi color-flux-tubes in terms of the dual Ginzburg
-Landau theory. We consider  two ideal  cases, where the directions of all 
the color-flux-tubes are  the same in one case and alternative 
in the other case for neighboring flux-tubes.
We formulate the system of multi color-flux-tubes
by regarding it as the system  of two color-flux-tubes 
penetrating through  a two dimensional  sphere surface.
We find the multi flux-tube configuration becomes uniform above  some 
critical flux-tube number density $\rho_c = 1.3 \sim 1.7 {\rm fm}^{-2}$.
On the other hand, the inhomogeneity on the color electric  distribution 
appears
when the flux-tube density is smaller than $\rho_c$.
We discuss the relation between the inhomogeneity in the color-electric 
distribution 
and  the flux-tube number density in the multi-flux-tube system
created during the QGP formation process in the 
ultra-relativistic heavy-ion collision. 
\end{abstract}

\newpage
\section{Introduction}

The QCD vacuum is non-trivial at zero temperature.
In this vacuum, quarks and gluons are confined in hadrons and the chiral 
symmetry is spontaneously broken.
As the temperature increases, however, the color
degrees of freedom in hadrons are defrozen. 
Above the critical temperature, the QCD vacuum is in the quark-gluon plasma
(QGP) phase, where quarks and gluons move almost freely.
The lattice QCD simulation demonstrates that such a phase transition 
happens at about 280 MeV for the pure gauge case\cite{karsch}
and at about 100$\sim$200MeV for the full QCD case
\cite{kanaya}.

In the recent years, some experimental groups are trying 
to create  QGP as the new form of matter 
in the laboratory using high-energy heavy-ion collisions.
The RHIC (Relativistic Heavy Ion Collider)  project is aimed at forming 
QGP and at studying its properties.
The scenario of producing QGP is based on Bjorken's picture\cite{bjorken}.
Just after  heavy ions pass through each other,
many color-flux-tubes are produced between the projectile and  the target,
and pulled by them as shown in Fig.1(a).
Usually, it is guessed that these flux-tubes are cut into several pieces 
through quark-antiquark pair creations, and
these short flux-tubes, which behave as excited 'mesons',
are thermalized by stochastic collisions among themselves.
If the energy deposition is larger than a critical value,
the thermalized system becomes the QGP phase, whereas if it is lower, 
the system remains to be the hadron phase.

The features of the multi color-flux-tube system strongly depend 
on their density of the flux-tubes created by  
hard process in early stage.
When the flux-tube number density is low enough, the system is 
approximated as 
the incoherent sum of the individual flux-tube.
Its evolution is expected to  be superposition of 
random  multiple hadron creations of many color-flux-tubes produced 
between many nucleon-nucleon pairs. 
On the other hand, when the flux-tube number density is  sufficiently high, many flux-tubes overlap each other 
and would be melted  into a big flux-tube.
During this process, each flux-tube loses  
its individuality  and the whole system can be regarded as a huge 
flux-tube  between heavy-ions like a condenser\cite{matsui}.

In this paper, we would like to study the  properties of the 
multi flux-tube system.
QCD is very hard to deal with in the infrared
region analytically due to the breakdown of the perturbation technique.
Moreover,
for such a large scale system
there is a severe limit on the computational power even in the 
lattice QCD simulation.
Hence, we resort to use  of the dual Ginzburg-Landau (DGL) theory
\cite{maedan,SST} for this subject.
This is the effective theory of nonperturbative QCD and can describe 
the color confinement, where there appears a long range
linear potential between a quark-antiquark 
pair\cite{suganuma}.
It is also able to describe the color-flux-tube as the dual version of
the Abrikosov vortex\cite{abrikosov}.

In Sect.2 we formulate the system of multi color-flux-tubes by regarding
it as that of two color-flux-tubes penetrating on a two dimensional sphere 
surface for neighboring flux-tubes.
In this paper we shall study the 
qualitative aspects of the multi flux-tube system.
We consider two cases; i.e. the directions of  the color-flux-tubes are 
the same or alternative.
The numerical results are discussed in Sect.3. Sect.4 is devoted to
the concluding remarks  and discussions on the QGP formation. 

\section{Multi color-flux-tube system in the dual Ginzburg-Landau theory}
In this section, we formulate the multi-color-flux system in the
dual Ginzburg-Landau (DGL) theory.
As 't Hooft proposed in 1981,  SU($N_c$) gauge 
theory is reduced to  U(1)$^{N-1}$ gauge theory with
the monopole  by abelian gauge fixing\cite{t Hooft}.
In this gauge, there appears a monopole as a topological object,
whose condensation leads to color confinement through the dual Meissner 
effect.  
Based on this idea, the DGL 
theory\cite{maedan,SST,ezawa,ichie} was proposed 
as an effective theory of the nonperturbative QCD.
The DGL lagrangian in the pure gauge system is written by using
the dual gauge field,
$B_\mu=\vec B_\mu \cdot \vec H = B_\mu^3 T^3+B_\mu^8 T^8$,
 and QCD-monopole field, $\chi=\sqrt{2} \sum_a    \chi_a E_a$ with 
$E_1=\frac{1}{\sqrt{2}}(T_6+iT_7)$, $E_2=\frac{1}{\sqrt{2}}(T_4-iT_5)$ and 
$E_3=\frac{1}{\sqrt{2}}(T_1+iT_2)$;
\begin{eqnarray}
{\cal L}_{DGL}
={\rm tr}{\hat {\cal L}} \nonumber 
\end{eqnarray}
\begin{eqnarray}
\hat {\cal L}=-{1 \over 2}
(\partial _\mu  B_\nu -\partial _\nu B_\mu )^2 +
[\hat{D}_\mu, \chi]^{\dag}[\hat{D}_\mu, \chi]
-\lambda ( \chi^{\dag} \chi -v^2)^2,
\label{eq:dgllag}
\end{eqnarray}
where $\hat{D}_\mu=\hat{\partial}_\mu +igB_\mu$ is the dual covariant derivative.
The second term leads the usual form as following,
\begin{eqnarray}
{\rm tr}([\hat{D}_\mu, \chi]^{\dag}[\hat{D}_\mu, \chi])
&=&
{\rm tr}([\hat{\partial}_\mu +igB_\mu, \chi^{\dag}]
[\hat{\partial}_\mu +igB_\mu, 
\chi]) \nonumber\\ 
&=&2{\rm tr}\{
(\partial_\mu \chi^*_a E_{-a} +i g\vec B_\mu \chi^*_a [\vec{H}, E_{-a}])
(\partial_\mu \chi_b E_b + i g\vec B_\mu \chi_b [\vec{H}, E_b])\}
\nonumber \\ 
&=&\left|(\partial_\mu+ig\vec \alpha_a \cdot \vec B_\mu )
\chi _a  \right|^2,
\end{eqnarray}
where $\alpha_a(a=1,2,3)$ are the root vectors of the SU(3) algebra,
$\vec \alpha_1=(-\frac12, \frac{\sqrt{3}}{2})$,
$\vec \alpha_2=(-\frac12, -\frac{\sqrt{3}}{2})$,
$\vec \alpha_3=(1,0)$.
The dual gauge field $B_\mu$ is defined on the dual gauge manifold
U(1)${}^m_3\times$U(1)${}^m_8$,
which is the dual space of the maximal torus subgroup
U(1)${}^e_3\times$U(1)${}^e_8$
embedded in the original gauge group SU(3).
The abelian field strength tensor is written as  
$F_{\mu\nu}=
{}^* (\partial\wedge B)_{\mu\nu}$
so that the role of the electric and the magnetic field are interchanged 
in comparison with the ordinary $A_\mu$  description.
The QCD-monopole has magnetic charge  $g\vec \alpha_a$,
where $g$ is the dual gauge coupling constant.
There appears the Dirac quantization condition, $eg = 4\pi$,
with the electric charge gauge coupling constant $e$.

We consider the flux-tube with a quark and an antiquark at the both ends. 
In the standard notation\cite{kerson,SST}, 
the quark charges  are $\vec Q_a \equiv e\vec w_a$,
where $\vec w_a(a=1,2,3)$ are the weight vectors of the SU(3) algebra,
$\vec w_1 = (\frac{1}{2}, \frac{1}{2\sqrt{3}})$,
$\vec w_2 = (-\frac{1}{2}, \frac{1}{2\sqrt{3}})$,
$\vec w_3 = (0,- \frac{1}{\sqrt{3}})$,
for the three color 
states\cite{kerson},
red($R$), blue($B$) and green($G$), respectively.
Using the Gauss law, one finds the color electric field $\vec{\bf{ E}}$
and then the dual gauge field $\vec B_\mu$, obeying 
$\vec F_{\mu\nu}={}^*(\partial \wedge \vec B)_{\mu\nu}$,
are proportional to  the color charge $\vec Q$.
The QCD-monopole $\chi_a$ couples with the quark charge $\vec Q_b$
in the form of $\vec \alpha_a \cdot \vec Q_b \chi_a$.
We note the algebraic relation between the root vector and the weight 
vector as,
\begin{eqnarray}
2\vec\alpha_a \cdot \vec w_b = \sum_{c=1}^3 \varepsilon_{abc}
 \in\{-1,0,1\}, 
\end{eqnarray}
and therefore  one kind of QCD-monopole $\chi_a$ couples with
not three but  two of the quark charges.
For the example, in the case of ($R$-$\bar R$) system,
$|\chi_1|$ is never affected; $|\chi_1| = v$.
For this case, we can rewrite Eq.(\ref{eq:dgllag}) as, 
\begin{eqnarray}
{\cal L}_{DGL}=-{1 \over 3}\cdot{1 \over 4}
(\partial _\mu B^{\rm R}_\nu -\partial _\nu B^{\rm R}_\mu )^2
+2\left| {(\partial _\mu +\frac{i}{2} gB^{\rm R}_\mu )\chi^{\rm R}} \right|^2
-2\lambda (\left| \chi ^{\rm R}  \right|^2-v^2)^2,
\end{eqnarray}
where $\vec B_\mu \equiv \vec w_1 B_\mu ^{\rm R} $ and 
$|\chi_1| = v$, $\chi_2=\chi^{{\rm R}*}$ and $\chi_3=\chi^{\rm R}$, because 
the system is invariant under the transformation,
$\chi_2^{{\rm R}*} \leftrightarrow                \chi_3^{\rm R}$. 
In this case, we can rewrite the DGL lagrangian in the simple GL form;
\begin{eqnarray}
{\cal L}_{DGL}=-{1 \over 4}
(\partial _\mu \hat B_\nu -\partial _\nu \hat B_\mu )^2
+\left| {(\partial _\mu +i\hat g\hat B_\mu )
\hat \chi _{}} \right|^2
-\hat \lambda (\left| {\hat \chi } \right|^2-\hat v^2)^2,
\end{eqnarray}
where the field variables and coupling constants are redefined as
\begin{eqnarray}
\hat B_\mu\equiv {1 \over {\sqrt 3}}B^{\rm R}_\mu,
 \hspace{0.5cm}
\hat g\equiv {{\sqrt 3} \over 2}g,
 \hspace{0.5cm}
\hat \chi \equiv \sqrt 2\chi^{\rm R},
 \hspace{0.5cm}
\hat v\equiv \sqrt 2v,
 \hspace{0.5cm}
\hat \lambda \equiv {\lambda \over 2}.
\end{eqnarray}
We get the same expression for the other two cases ($B$-$\bar B$,
$G$-$\bar G$).
Hereafter, we will drop the notation $\hat{}$ since there is no confusion.
The field equations for $B_\mu$  and $\chi$ are derived by the 
extreme condition,
\begin{eqnarray}
\partial ^2\chi +2igB^\mu (\partial _\mu \chi )
+ig(\partial _\mu B^\mu )\chi
-g^2B_\mu^2\chi +2\lambda (|\chi |^2-v^2)^2\chi =0,
\end{eqnarray}
\begin{eqnarray}
\partial _\mu (\partial ^\mu B^\nu
-\partial ^\nu B^\mu )
+ig\{(\partial ^\nu \chi ^*)\chi
-(\partial ^\nu \chi )\chi ^*\}
+2g^2B^\nu |\chi |^2=0.
\end{eqnarray}

We consider two ideal cases of multi color-flux-tube
penetrating on a {\it two dimensional  plane}.
The directions of all the color-flux-tubes are the same[Fig.2a] in one 
case or alternative[Fig.2b] in the other case.
When the flux-tubes are long enough,
the effect of  the flux edges is negligible.
Hence, taking the direction of the flux-tubes as the $z$-axis, the system 
is translationally invariant in the $z$-direction and 
is essentially described only with two spatial coordinates $(x,y)$.
For the periodic case in the $(x,y)$ coordinate, 
we can  regard the system as two flux-tubes going through two poles 
(north and south poles) of the $S^2$ {\it sphere}.
For the  system of  flux-tubes with all the directions being the same,
we take two flux-tubes passing through the two poles on $S^2$ sphere
as shown in Fig.2(a) (which we call the two flux-tubes system).
For the alternative case, on the other hand,
we take a flux-tube coming in from the south pole and the other leaving 
out 
through the  north pole (which we call
flux-tube and anti-flux-tube system).
Such a prescription leads the exact solution for the periodic crystal of 
the sine-Gordon kinks, and also provides a simple but good description
for the finite density Skyrmion system studied by 
Manton\cite{sine,skyrmion}.

The two color-flux-tube system on the sphere $S^2$  with 
radius $R$  corresponds to the multi-flux-tube system with the 
density $\rho=1/(2\pi R^2)$.
Introducing the polar coordinates $(R,\theta,\varphi)$ on $S^2$,
we consider the  static solution satisfying 
\begin{eqnarray}
B_0=0 , \hspace{2cm} {\bf B}=B(\theta){\bf e}_\varphi
\equiv \frac{\tilde B(\theta)}{R\sin \theta}  {\bf e}_\varphi ,
\hspace{2cm} \chi = \bar \chi(\theta) e^{in\varphi},
\end{eqnarray}
where we have used the axial symmetry of the system.
Here the electric field penetrates vertically on the {\it sphere} {surface,
${\bf E}//{\bf e}_r$;
\begin{eqnarray}
{\bf E}={\bf \nabla} \times {\bf B} = E{\bf e}_r,
\end{eqnarray}
which corresponds to the electric field penetrating vertically on the {\it 
plane}, ${\bf E}//{\bf e}_z$.
The field equations  are given by
\begin{eqnarray}
{1 \over {R^2\sin \theta }}
{d  \over {d \theta }}
(\sin \theta {{d \bar \chi }
\over {d \theta }})
-[{1 \over {R^2\sin ^2 \theta }}(n-g \tilde B(\theta))^2
+2\lambda (\bar \chi ^2-v^2)]\bar  \chi =0,
 \label{eq:ren1}
\end{eqnarray}
\begin{eqnarray}
{d  \over {R^2d \theta }}
({1 \over {\sin \theta }}
{d  \over { d \theta }} \tilde B(\theta))
+\frac{2g}{\sin \theta}(n-g\tilde B(\theta))\bar \chi ^2=0.
\label{eq:ren2}
\end{eqnarray}
Consider the closed loop $C$ on $S^2$ with a constant  
 polar angle  $\theta = \alpha$ and $\phi \in [0,2\pi)$,
the electric flux penetrating the area surrounded by the loop $C$ is given 
by
\begin{eqnarray}
\Phi( \alpha ) = \int_{S} {\bf E} \cdot d{\bf S}
  = \int {\bf \nabla} \times {\bf B} \cdot
    d{\bf S}
  = \oint_c {\bf B} \cdot d{\bf l}
  = 2 \pi \tilde B(\alpha).
\end{eqnarray}
The boundary conditions for the two flux-tubes  system 
as shown in Fig.2(a) are given as 
\begin{eqnarray}
\Phi( \alpha )=2 \pi \tilde B(\alpha) = 0,
  \hspace{2cm}              \bar \chi(\alpha) = 0   \hspace{2cm}
{\rm as}  \hspace{0.5cm} \alpha  \rightarrow 0,  
\label{eq:kashi1}
\end{eqnarray}
\begin{eqnarray}
\Phi( \alpha )=2 \pi  \tilde B(\alpha)
= \frac 12  \int _{S} E R^2 d\Omega 
= \pm \frac{2\pi n}{g}
\hspace{2cm} {\rm as} \hspace{0.5cm}
\alpha \rightarrow \frac{\pi}{2} \pm \epsilon,
\label{eq:kashi2}
\end{eqnarray}
Here, $n$ corresponds to the topological number of the flux-tube,
which appears also in the vortex solution in the superconductivity.
This boundary condition at $\alpha \rightarrow \frac{\pi}{2} \pm \epsilon$
has a discontinuity for the dual gauge field $\tilde B$.
Because the electric flux leaves out from the two poles,
there should be some sources to provide the electric flux.
In this case,
the Dirac-string like
 singularity appears on  the $\theta=\pi/2$ line, through which the electric
flux comes into the sphere from long distance. 
For the flux-tube and anti-flux-tube system as shown in Fig.2(b), the 
boundary condition around $\theta = 0,\pi$ is given as  
\begin{eqnarray}
\Phi( \alpha )=2 \pi \tilde B(\alpha) = 0,
  \hspace{2cm}              \bar \chi(\alpha) = 0   \hspace{2cm}
{\rm as}  \hspace{0.5cm} \alpha  \rightarrow 0,\pi.  
\end{eqnarray}
In this case, there does not appear the Dirac-string like singularity,
since the electric flux is conserved.
The free energy for the unit length
of the color-flux-tube is written as
\begin{eqnarray}
F&=&\int { R^2 d\Omega}
\left[ {{1 \over 2}\left( {{1 \over {R^2\sin \theta }}
{d \over {d\theta }}\tilde B(\theta )} \right)^2
+\left( {{1 \over R}{{d\bar \chi } \over {d\theta }}} \right)^2} \right] 
\nonumber \\
&+&
\int { R^2 d\Omega}\left[ {{1 \over {R^2\sin ^2\theta }}
(n-g\tilde B(\theta ))^2\bar \chi ^2+\lambda
 \left( {\bar \chi ^2-v^2} \right)^2} \right] \nonumber\\
&=&\int_0^{\theta =\pi } 
{2\pi\sin }\theta d\theta \left[ 
{\left( {{{d\bar \chi } \over {d\theta }}} \right)^2
+{1 \over {\sin ^2\theta }}(n-g\tilde B(\theta ))\bar \chi ^2} 
\right]\nonumber\\
&+&\int_0^{\theta =\pi } {2\pi\sin }\theta d\theta \left[ 
{{1 \over 2}\left( {{1 \over {\sin \theta }}
{d \over {d\theta }}\tilde B(\theta )} \right)^2}
 \right]\cdot {1 \over {R^2}}+\int_0^{\theta =\pi } 
 {2\pi\sin }\theta d\theta 
 \left[ {\lambda \left( {\bar \chi ^2-v^2} \right)^2} \right]\cdot R^2.
\end{eqnarray}
First, we consider a limit of $R \rightarrow \infty$, which corresponds
to the ordinary single vortex solution.
Introducing a new variable $\rho \equiv R \sin \theta$,
the free energy in  the limit $R \gg \rho$$(\theta \sim 0)$ is written as
\begin{eqnarray}
F&=&\int_{}^{} {2\pi \rho d\rho}\left[
 {{1 \over 2}\left( {{1 \over \rho }
 {d \over {d\rho }}\rho B(\rho )} \right)^2
 +\left( {{{d\bar \chi } \over {d\rho }}} \right)^2} \right] \nonumber \\
&+&\int_{}^{} {2\pi \rho d\rho}\left[ {{1 \over {\rho ^2}}
(n-g\rho B(\rho ))\bar \chi ^2
+\lambda \left( {\bar \chi ^2-v^2} \right)^2} \right],
\label{eq:enchu1}
\end{eqnarray}
and the  field equations are
\begin{eqnarray}
{1 \over \rho }{d \over {d\rho }}\rho \left( {{d \over {d\rho }}\bar  \chi } 
\right)
-{1 \over {\rho ^2}}(n-g\rho B(\rho ))^2
-2\lambda \left( {\bar \chi ^2-v^2} \right)\bar \chi =0,
\end{eqnarray}
\begin{eqnarray}
{d \over {d\rho }}{1 \over \rho }
{d \over {d\rho }}\left( {\rho B(\rho )} \right)
+{{2g} \over {\rho ^{}}}(n-g\rho B(\rho ))\bar \chi ^2=0,
\label{eq:enchu2}
\end{eqnarray}
with the boundary condition,
\begin{eqnarray}
\Phi = 2\pi\rho B(\rho)|^{\infty}_0=\frac{2\pi n}{g}
\hspace{0.5cm} \mbox{as} \hspace{0.5cm} \rho \rightarrow \infty.
\end{eqnarray}
Above equations, (\ref{eq:enchu1}-\ref{eq:enchu2}),
coincide exactly with those of ordinary  single vortex 
solution in the cylindrical coordinate.
Thus, we get the desired results.

One can analytically investigate the dependence of the  profile 
functions$(\tilde B(\theta),\chi(\theta))$ on the flux-tube number density.
For this purpose, we express the free energy as
\begin{eqnarray}
F \equiv  f_0+f_E\cdot {1 \over {R^2}}
+f_\chi \cdot R^2,
\end{eqnarray}  
where $f_0,f_E,$ and$f_\chi$ are $R$ independent functions and written as 
\begin{eqnarray}
f_0  & \equiv & 
\int_0^{\theta =\pi } 
{2\pi\sin }\theta d\theta \left[ {\left( {{{d\bar \chi } \over {d\theta }}} 
\right)^2
+{1 \over {\sin ^2\theta }}(n-g\tilde B(\theta ))^2\bar \chi ^2} \right], \\
f_E & \equiv &
\int_0^{\theta =\pi }
 {2\pi\sin }\theta d\theta  {{1 \over 2}
 \left( {{1 \over {\sin \theta }}{d \over {d\theta }}
 \tilde B(\theta )} \right)^2},
\\
 f_\chi & \equiv &
\int_0^{\theta =\pi }
 {2\pi\sin }\theta d\theta  {\lambda 
 \left( {\bar \chi ^2-v^2} \right)^2} .
\end{eqnarray}
In the large  $R$ case, which corresponds to the small 
color-flux-tube 
number density in the original multi-flux-tube system, the third term 
$f_\chi R^2$ is dominant.
Hence, the free energy $F$ is minimized as $\bar \chi \sim v$,
that is, the QCD-monopole tends to condense, and then the color electric 
field is localized only around $\theta=0$ (north pole) and $\theta=\pi$
(south pole) as shown in Fig.3.
On the other hand, in the small  $R$ case,
the second term ${f_E}/{R^2} $ is dominant.
There is a constraint on the total flux penetrating on the upper sphere,
\begin{eqnarray}  
\Phi \equiv \int^{\frac{\pi}{2}}_0 E(\theta)2 \pi R^2 {\rm sin}
\theta d \theta = \frac{2 n\pi}{g},
\end{eqnarray}
that is,
\begin{eqnarray}
\int^1_0 E(t) dt= \frac{n}{gR^2} \equiv C \hspace{0.5cm}
 \mbox{with} \hspace{0.5cm} t = {\cos}\theta.  
\end{eqnarray}
Hence, one finds the equation,
\begin{eqnarray}
f_E \propto \int^{\frac{\pi}{2}}_{0}E(\theta)^2 {\rm \sin}\theta d \theta
=\int^1_0  E(t)^2 dt
=\int^1_0 \{(E(t)-C)^2\}dt +  C^2  
\geq C^2.
\end{eqnarray}
This condition leads to the  uniform color electric field $E = C$,
which provides the minimum of $f_E$.
Thus, the color electric field tends to spread over the space uniformly.

Finally we consider the critical radius $R_c$ of the phase transition to 
the 
normal phase, where the QCD-monopole disappears.
There are three useful inequalities on $f_E$, $f_\chi$, and $F$,
\begin{eqnarray}
f_E \geq 2\pi (\frac{ n}{g})^2
\label{eq:shikic}
\end{eqnarray}
\begin{eqnarray}
0 \leq f_\chi \leq\ 4 \pi \lambda v^4, \label{eq:shikia}
\end{eqnarray}
\begin{eqnarray}
F=f_0+f_E \cdot \frac{1}{R^2} + f_\chi \cdot R^2 \geq f_0 +2\sqrt{f_E 
f_\chi}.
\label{eq:shikib}
\end{eqnarray}
The equality is satisfied in Eq.(\ref{eq:shikib}),
\begin{eqnarray}
R^4 = \frac{f_E}{f_\chi} 
\end{eqnarray}
Using the inequalities equations(\ref{eq:shikic}-\ref{eq:shikib}),
$R^4$ is larger than  the critical $R^4_c$;
\begin{eqnarray}
R^4 \geq R^4_c\equiv
\frac{2\pi n^2/g^2}{4 \pi \lambda v^4 }=
\large( \frac{n^2}{2\lambda g^2}\large)\frac{1}{v^4}.
\end{eqnarray}
For $R > R_c$, there exists a nontrivial inhomogeneous solution, which
differs from the normal phase.
For $R \leq R_c$, homogeneous normal phase provides the minimum of $F$.
Thus, $R_c $
 is the critical radius of the phase transition from the flux-tube
 phase to the normal one. 
In this case, the critical radius and the electric field are given by 
\begin{eqnarray}
\rho_c=\frac{1}{2 \pi R^2_c }=
\sqrt{\frac{\lambda}{2}}\frac{gv^2}{\pi n},
\label{eq:analy} 
\end{eqnarray}
\begin{eqnarray}
E_c
=\frac{n}{gR_c^2}=\sqrt{2\lambda}v^2, 
\end{eqnarray}
respectively.

\section{Numerical Results}
We start with showing the low density case of two flux tubes system in Fig.3.
The parameters of the DGL theory are fixed as $\lambda=$ 25,
$v=$ 0.126GeV, which lead the masses $m_B=$ 0.5GeV and 
$m_\chi=$ 1.26GeV\cite{SST}.
This parameter set provides the flux-tube radius $r_{FT}\sim$0.4fm and the 
suitable  interquark potential with 
the string tension as $\sigma=$1GeV/fm.
 The QCD-monopole condensate $\bar \chi(\theta)$,
the  dual gauge field $\tilde B(\theta)$ and
the color electric field  $E(\theta)$
are plotted as  functions of 
the polar angle $\theta$.
The electric field $E(\theta)$ is localized around 
the two poles($\theta$=0 and $\pi$)
and drops suddenly as $\theta$ deviates from the two poles.
The QCD-monopole condensate  vanishes at the two poles and 
becomes constant in the region away from these poles.
This behavior corresponds to the case of independent two vortices
in superconductivity.
Different from these physical quantities, the dual gauge field $\tilde 
B(\theta)$ is discontinuous at $\theta=\pi/2$.
There should be Dirac-string like  source to provide,
because the electric flux leaving out from the two poles.
It should be noted that the system has the reflection symmetry on
$\theta=\frac{\pi}{2}$ plane.

We show now in Fig.4 the number density dependence of the flux-tubes.
For the large radius of the sphere, ($R \geq $2fm), the color electric flux
is localized at $\theta=0$ and $\pi$ and there 
the  QCD-monopole condensate vanishes, 
while  
the $\bar \chi$  becomes constant $\bar \chi \simeq v$ around 
$\theta = \frac{\pi}{2}$.
As the radius $R$ decreases, the electric flux, localized at $\theta=0$,
$\theta=\pi$, starts to  overlap, and the value of the QCD-monopole 
condensate $\bar \chi$ becomes small.
The electric field $E(\theta)$ 
becomes constant and
$\bar \chi$ vanishes below  a critical radius $R_c
$.
We show in Fig.5 the QCD-monopole condensate at $\theta=\frac{\pi}{2}$
(the maximum value of the QCD-monopole condensate) as a function of the 
sphere radius $R$ (flux-tube number density $\rho=1/(2 \pi R^2)$).
The (first order) phase transition
occurs and the system becomes homogeneous normal 
phase above the critical value of the flux-tube number density.

Here, we compare the free energy of two flux tube system with
that of inhomogeneous system in Fig.6.
At large $R$, the former is smaller and the system favors the
existence of two flux tubes.
As  $R$ becomes smaller,
the energy difference  of the system becomes smaller.
Two flux-tubes melt and the electric field are changed to be 
homogeneous below the critical radius $R_c=0.35$fm, which   
 corresponds to the critical density $\rho_c=1/(2 \pi R_c^2)=
 1.3{\rm fm}^{-2}$. 
This critical density agrees  with the 
analytic estimation in Eq.(\ref{eq:analy}),
$\rho_c=
\sqrt{\frac{\lambda}{2}}\frac{gv^2}{\pi n}=
1.3{\rm fm}^{-2}$.  

We discuss now the system of flux-tube and anti-flux-tube with opposite 
direction placed at $\theta=$0 and $\theta=\pi$ respectively 
as shown in Fig.7.
At low flux-tube number density ( $R \geq 2$fm),
the flux-tube  is localized at $\theta=0$, while the anti-flux-tube  at 
$\theta=\pi$.
The QCD-monopole condensate $\bar \chi(\theta)$ vanishes at the 
two poles ($\theta=0,\pi$) and becomes constant
away from these poles.
As  $R$ decreases, the electric flux starts to cancel each other 
and the QCD-monopole
condensate becomes small.
We also compare  the free-energy of the flux-tube and anti-flux-tube system
with the homogeneous system, in which both QCD-monopole condensate
and electric field are vanished. The critical radius, $R_c=$ 0.31fm,
is similar to the value of the two flux-tubes system.

\section{Summary and Concluding Remarks}

We have studied the system of multi color-flux-tubes using  the dual 
Ginzburg-Landau (DGL) theory.
The DGL theory provides the color-flux-tube between a quark and 
an anti-quark pair as the dual version of the Abrikosov vortex.
We have considered two ideal  cases, where the directions of all the 
color-flux-tubes are the same and alternative penetrating
on a two-dimensional plane.
In order to treat these cases in a simple way, we have introduced 
color-flux-tubes on a 
sphere $S^2$ going through its north and south poles.
Here, the flux-tube system on the sphere $S^2$ with the radius $R$
corresponds to the multi-flux-tube system with the flux-tube number density 
$\rho=1/(2\pi R^2)$.
The energy minimization condition of the system provides 
a simple coupled diffential equation for the dual gauge field 
$\tilde B(\theta)$ and QCD-monopole
condensate $\bar \chi(\theta)$ as functions of the polar angle $\theta$.
We have solved the coupled equation numerically
with the parameter set, which provides the  radius of flux 
tube as $r_{FT}=0.4$fm in the flat space ($R \rightarrow \infty$).

We have found in both cases that the solution behaves as two independent
flux-tubes at a large $R$, i.e., a small number density $\rho$.
As the radius $R$ decreases, 
the QCD-monopole condensate decreases 
and eventually vanishes at a critical density,
where the color-electric field $E$ becomes uniform.
The critical density $\rho_c$ is found as 
$\rho_c=$1.3${\rm fm}^{-2}=(1.14{\rm fm})^{-2}$ 
for the two color-flux tube system;
$\rho_c=$1.7${\rm fm}^{-2}=(1.3{\rm fm})^{-2}$
for the  flux-tube and anti-flux-tube system.
Such similar values   in both cases suggest  that an actual 
flux-tube system would become uniform  around similar
density  to  $\rho_c \sim$1.5fm${}^{-2}$,
because realistic flux-tube system includes the flux-tubes and anti-flux 
tubes randomly, which  would correspond to an intermediate system between 
the above two ideal cases.  
Thus, the configuration of the color-electric field and the QCD-monopole field 
depend largely on the flux-tube density.

As discussed above, many flux-tubes are melted 
around $\rho_c \sim$ 1.5 fm${}^{-2}$.
Let us discuss here in  case of central collision between
heavy-ions with mass number $A$.
The nuclear radius $R$ is given by $R=R_0 A^{1/3}$, where
$R_0$=1.2fm corresponds to the nuclear radius,
and the normal baryon-number density is $\frac{A}{\frac{4}{3}\pi R^3}
=\frac{1}{\frac43 \pi R_0^3}$.
In the central region, one nucleon in the projectile makes hard
collisions with $\frac32 A^{1/3}$$(=2 \pi R_0^3 A^{1/3} \times \rho_0)$
nucleons in the target,
because the reaction volume is $\pi R_0^2 \times 2 R =2\pi R^3_0 A^{1/3}$.
Hence, nucleon-nucleon collision number is expected as 
 $(\frac32 A^{1/3})^2= \frac{9}{4} A^{2/3}$ 
per the single-nucleon area $\pi R_0^2$ between projectile and target 
nuclei.
Assuming one flux-tube formed in one nucleon-nucleon hard collision,
the flux-tube density is estimated as 
$\rho=\frac{ \frac{9}{4} A^{2/3}}{\pi R_0^2}=\frac{A^{2/3}}{(1.4{\rm fm})^2}$.
For instance,
$\rho$ would be 4.5, 5.8 and 17.5 fm${}^{-2}$ for $A=$ 27, 40 and 208.
This would indicate that the scenario of creating large sizes of flux tube 
becomes  much relevant for $A$-$A$ collision with larger $A$.

It would be important to reconsider  the process of the QGP formation 
in terms of the flux-tube number density.
When the flux-tube density is low enough, the flux-tubes are localized.
Each flux-tube evolution would be regarded as the multi creation of hadrons
in the high energy p-p collision via the flux-tube breaking.
In this process, $q$-$\bar q$ pair creation plays an essential role 
on the QGP formation, which is the usual scenario.

On the other hand, for the dense flux-tube system,
neighboring flux-tubes are melted into a large cylindrical tube,  
where QCD-monopole condensate disappears.
Such a system, where the color electric field is made between heavy-ions,
becomes approximately homogeneous and is regarded as the 
'color condenser'.
In this case, large homogeneous QGP may be created in the central region.

In the actual case, however, the variations and directions 
of the color-flux-tubes are random.
For instance, in the peripheral region,
flux-tubes would be localized and are broken by 
quark-antiquark pair creations. 
In the central region,  dense flux-tubes are melted by
 annihilation or unification\cite{ichie2} of flux-tubes.
 In this region, a huge system of dynamical gluons appears,
 because many dynamical gluons are created during this process.
Thermalization of such quarks and gluons leads to   QGP.  
Thus, the process  of QGP formation should  depend largely on the 
density of created flux-tubes, 
which is closely related to the incident energy, the impact parameter and 
the size of the projectile and the target nuclei.

\newpage

Figure caption

\noindent Fig.1 The scenario of the QGP formation in high energy heavy ion collisions.
(a) There appear many color-flux-tubes between the projectile  nucleus and 
the target nucleus just after the collisions.
(b) When the distance between the two nuclei becomes large, there appears 
the pair creation of quark and anti-quark.
(c) Many created quarks and anti-quarks make 
frequent collisions to form a thermal equilibrium and form QGP when the 
energy density is larger than the critical value.

\noindent Fig.2 (a) A multi color-flux system with the same direction of the 
flux-tubes is approximated by the two  color-flux-tubes going out 
from the north and the south poles on a sphere $S^2$.
(b) A multi color-flux system, where flux-tube direction is alternative, is 
approximated by the flux-tube and  'anti-flux-tube' system penetrating on 
$S^2$
with the color-flux going in from the south pole and leaving
out from the north pole.

\noindent Fig.3 The color-electric field $E(\theta)$(solid curve), 
the QCD-monopole condensate $\bar \chi(\theta)$(dashed curve)
and the dual gauge field $B(\theta)$ (thin solid curve) 
are plotted as  functions of 
the polar angle $\theta$ for the low density case; $R=$4.0fm.

\noindent Fig.4 We show the case of the two flux-tubes system on penetrating $S^2$.
The color-flux $E(\theta)$ and the QCD-monopole condensate $\bar \chi(\theta)$
are depicted as  functions of the polar angle 
$\theta$ for the three radii; $R=$2.0fm, 0.5fm and 0.347fm.
Below the critical radius $R_c$=0.347fm, the color electric field 
E  becomes constant and the QCD-monopole condensate $\bar \chi$ 
vanishes entirely.

\noindent Fig.5 We show the $R$ (flux-tube density, $1/(2\pi R^2)$) dependence
of QCD-monopole configuration.
The QCD-monopole condensate at $\theta=\frac{\pi}{2}$  decreases,
as the radius $R$ becomes smaller, and vanishes below $R=R_c$.

\noindent Fig.6 We show the free energy in two color-flux-tubes system
(solid curve) and
uniform system (dashed curve).
Because the lower free energy system is realized,
the inhomogeneous system  is changed into homogeneous system
below the critical radius, $R_c=0.35{\rm fm}$ 
($\rho_c = 1/(2\pi R_c^2) = 1.3{\rm fm}^{-2}$).

\noindent Fig.7 We show the case of the flux the and anti-flux-tube penetrating on $S^2$
system. 
The color-electric field $E(\theta)$
and the QCD-monopole condensate  $\bar \chi(\theta)$ are depicted  as  functions 
of 
the 
polar angle $\theta$ for the three radii; $R=$2.0fm, 0.5fm, and  0.315fm.
Below $R=$0.315fm, both the color electric flux and the QCD-monopole 
condensate vanish entirely.


\begin{thebibliography}{99}
\bibitem{karsch}
G.~Boyd, J.~ Engels, F.~Karsch, E.~Laermann, 
C.~Legeland,  M.~Lutgemeier and B.~Petersson,
Phys.~Rev.~Lett.~{\bf 75} (1995) 4169. \\
J.~Fingberg, F.~Karsch and  U.~M.~Heller, Nucl.~Phys. (Proc.
Suppl.) ~{\bf B}{\bf 30} (1993) 343.
     \bibitem{kanaya}
K.~Kanaya, Prog.~Theor.~Phys.~(Suppl.) {\bf 120} (1995) 25.\\
Y.~Iwasaki, K.~Kanaya, S.~Sakai, T.~Yoshie,
Nucl.~Phys. (Proc. Suppl.) {\bf B26} (1992) 311.  
     \bibitem{bjorken}
J.~D.~Bjorken, Phys. Rev. {\bf D27} (1983) 140.
     \bibitem{matsui}
N.~K.~Glendenning and T.~Matsui, 
Phys.~Rev.~{\bf D28} (1983) 2890; \\
Phys.~Lett.~{\bf B141} (1984) 419.
     \bibitem{maedan}
T.~Suzuki, Prog.~Theor.~Phys. {\bf 80} (1988) 929.\\ 
S.~Maedan and T.~Suzuki, Prog.~Theor.~Phys. {\bf 81} (1989) 229; 
{\bf 81} (1989) 752.
     \bibitem{SST}
H.~Suganuma, S. Sasaki and H. Toki,
Nucl.~Phys.~ {\bf B435} (1995) 207. \\ 
H.~Suganuma, H.~Ichie, S.~Sasaki and H.~Toki, Proc.~of~Int.~Workshop
on ``Color Confinement and Hadrons'', (World Scientific, 1995) 65.
     \bibitem{suganuma}
H.~Suganuma and T.~Tatsumi, Phys.~Lett.
 {\bf B269} (1991) 371. \\ 
H.~Suganuma and T.~Tatsumi,
 Prog.~Theor.~Phys. {\bf  90} (1993) 379.
     \bibitem{abrikosov}
H.~B.~Nielsen and P.~Olesen, Nucl.~Phys.~{\bf B61} (1973) 45.
     \bibitem{t Hooft} 
G.~'t~Hooft, Nucl.~Phys.~{\bf B190} (1981) 455.
     \bibitem{ezawa}
Z.~F.~Ezawa and A.~Iwazaki,
Phys.~Rev.~{\bf D25} (1982) 2681.\\
Z.~F.~Ezawa and A.~Iwazaki,
Phys.~Rev.~{\bf D26} (1982) 631.
     \bibitem{ichie}
H.~Ichie, H.~Suganuma and H.~Toki, 
Phys.~Rev.{\bf ~D52} (1995) 2944 . \\
H.~Ichie, H.~Monden, H.~Suganuma and H.~Toki, 
Proc. of Int. Symp. on ``Nuclear Reaction Dynamics of Nucleon-Hadron
Manybody System'', Dec.~1995, RCNP~Osaka.
     \bibitem{kerson}
K. Huang, ``Quarks Leptons and Gauge Field'',~(World Scientific, 1982) 1.
     \bibitem{sine}
T.~D.~Lee, ``Particle Physics and Introduction to Field Theory''
(Harwood Academic Publishers, 1926) 1. 
     \bibitem{skyrmion}
N.~S.~Manton, Commun. Math.~ Phys. {\bf 111} (1987) 469.
     \bibitem{ichie2}
H.~Suganuma, S.~Sasaki, H.~Toki and H.~Ichie, Prog.~Theor.~Phys. 
(Suppl.) {\bf 120} (1995) 57.  
\end{thebibliography}
\end{document}